\documentstyle[aps,graphicx]{revtex}
\newcommand{\be}{\begin{equation}}
\newcommand{\ee}{\end{equation}}
\newcommand{\bea}{\begin{eqnarray}}
\newcommand{\eea}{\end{eqnarray}}
\newcommand{\bd}{\begin{displaymath}}
\newcommand{\ed}{\end{displaymath}}
\newcommand{\bi}{\begin{itemize}}
\newcommand{\ei}{\end{itemize}}
\newcommand{\bc}{\begin{center}}
\newcommand{\ec}{\end{center}}
\newcommand{\bfl}{\begin{flushleft}}
\newcommand{\efl}{\end{flushleft}}
\newcommand{\bfr}{\begin{flushright}}
\newcommand{\efr}{\end{flushright}}
\newcommand{\f}{\frac}

   \def\vk{\vec{k}}

\def\ra{\rightarrow}
\def\6{\partial} \def\a{\alpha} \def\b{\beta}
   \def\e{\epsilon}
  
  \def\l{\lambda}

\def\o{\omega}  \def\D{\Delta}
 \def\L{\Lambda}

\def\={\!\!\!&=&\!\!\!}
\def\+{\!\!\!&&\!\!\!+~}
\def\-{\!\!\!&&\!\!\!-~}

\begin{document}
\title{Some sum rules for non-Fermi Liquids: new applications taking into 
account the mass renormalization factor}
\author{\it J.J. Rodr\'{\i}guez--N\'u\~nez}
\address{Departamento de F\'{\i}sica, 
Funda\c{c}\~ao Universidade Federal do Rio Grande,   
96201-900 Rio Grande/RS, 
Brazil. \\e-m: jjrn@einstein.dfis.furg.br}
\author{\it I.\ \c{T}ifrea\footnote{Permanent address: 
Department of Theoretical Physics, University of Cluj, 3400 Cluj, Romania}}
\address{Universit\'a degli studi di Camerino,  
Dipartamento di Matematica e Fisica, 
via Madonna Delle Carceri, 62032 Camerino, Italy. \\e-m: 
tifrea@str2.unicam.it}
\author{\it S.\ G.\ Magalh\~aes}
\address{Departamento de Matem\'atica--CCNE, 
Universidade Federal de Santa Maria,   
97105-900 Santa Maria/RS, 
Brazil. \\e-m: ggarcia@ccne.ufsm.br}

\date{\today}
\maketitle
\begin{abstract}
Re-studying the non-Fermi liquid one--particle Green 
functions ($NFLGF$) we have extended the work of 
A.\ Balatsky (Phil.\ Mag.\ 
Lett.\ {\bf 68}, 251 (1993)) and L. Yin and S. Chakravarty 
(Int. J. Mod.\ Phys. B {\bf 10}, 805 (1996)), among others.  
We use the moment approach of W. Nolting (Z. Phys. {\bf 255}, 
25 (1972)) to compute the unknown parameters of the  
$NFLGF$'s in the framework of the Hubbard model. The zeroth 
order moment requires that our one--particle Green functions 
describe fermionic degrees of freedom. In order 
to satisfy the first order sum rule a 
renormalization, $\gamma \neq 1$, of the free electron mass 
is called for. The second order sum rule or moment imposes a 
relation between the {\it non--Fermi liquid} parameter, 
$\alpha$, the Coulomb interaction, $U$, and the frequency cutoff, 
$\omega_c$. We have calculated the effect of the mass 
renormalization factor, $\gamma$, on some physical quantities, like: 
1) the correlated momentum distribution function, $n_c(\vec{k})$, 
close to the effective chemical potential, at $T~=~0$; 2) the superconducting 
critical temperature, $T_c$; and 3) the superconducting critical 
interaction, $\lambda_{cr}$, and compare them with analytical results found 
in the literature. Also, we have calculated, for the first 
time, the isotope effect, $\alpha'$, for non--Fermi liquid systems, 
which reduces to $\alpha' = 1/2$ (the $BCS$ result) when 
$\alpha \rightarrow 0$. As a new case of non--Fermi liquid systems, 
in Appendix \ref{a20}, we have studied two inequivalent coupled 
Hubbard layers ($ICHL$) 
for which we calculate the one--particle spectral functions 
on the layers and 
perpendicular to them. We discuss the new features which appear 
due to the shift in the two effective chemical potentials and propose some 
experiments to detect the features found from our expressions.
\\
\\
Pacs numbers: 74.20.-Fg, 74.10.-z, 74.60.-w, 74.72.-h
\end{abstract}


\section{Introduction}\label{I}
\indent The unusual normal state properties in high temperature 
superconductors ($HTSC$)\cite{BM} have lead to the idea that the usual 
Fermi liquid theory is non longer valid. As a consequence of this,  
several phenomenological models\cite{Varma,Phil} have emerged with the 
purpose of explaining the strange metallic behaviour of the 
normal state phase and the behaviour of $T_c$ vs doping in the 
$HTSC$. We should like to point out that the origins of a 
non--Fermi liquid ground state for a strongly correlated 
material ($U \geq 2D$, where $U$ is the local repulsive 
Coulomb interaction and $2D$ is the width of the free 
band) in dimensions higher than one, namely, $d > 1$, is 
an issue that has not been solved analytically up to the 
present moment. The main arguments which have been cited as 
responsible for the failure of Landau theory are:
\begin{enumerate}
\item The high critical superconducting temperature is attributed to 
the $CuO_2$ planes, but is well known that in one or two dimensional 
systems superconductivity (or any spontaneous symmetry breaking) is 
suppressed due to fluctuation effects for $T \neq 0$;
\item The superconducting phase is very close to the long range 
magnetically ordered phase, and the exchange interaction $J$ in this 
strange metal can be too strong;
\item The concentration of holes $x$, or the carrier number, is too 
low;
\item The Coulomb interaction may be too strong so the adiabatic 
assumption in Landau's theory may no apply;
\item For optimally doped materials, the electrical resistivity
on the plane, $\rho_{ab} \propto T$ down to $T_c$. For a Fermi 
liquid, $\rho_{ab} \propto T^2$. There is not 
consensus regarding the origin of the anomalous in--plane 
transport. Two broad classes of theories attribute the anomalous behaviour  
either to singular forward or large--momentum scattering\cite{Hlubina}. 
Furthermore, Boltzmann transport equation gives an expression for the 
magnitude of the resistivity in terms of band parameters and a 
mean--free path between quasi--particle collisions. At low temperatures 
this expression suggests a mean--free path which is much larger that 
the lattice constant, as in conventional metals. However, at higher 
temperatures the resistivity smoothly increases to large values, 
suggesting a mean--free path which is smaller than the lattice 
constant, implying the breakdown of a 
quasi--particle picture\cite{Merino}.
\item The overdoped materials exhibit a ratio of resistivities, 
$\rho_{c}/\rho_{ab} \neq f(T)$ as in standard anisotropic materials. 
On the other hand, underdoped materials exhibit a divergent 
out--of--plane $\rho_c$ as $T$ decreases, even if $\rho_{ab}$ 
is metallic. The c-axis data are not as universal between different 
cuprate families. 
\item The Hall coefficient, $R_H(T)$, in $HTSC$ in the normal 
state shows the striking $NFL$ behaviour\cite{Kontani}. 
$R_H(T)$ follows a 
Curie--Weiss type $T$--dependence and $|R_H(T)| >> 1/|n~e|$ for 
$T \rightarrow 0$ in the underdoped compounds, where 
$n$ is the carrier number.  Moreover, $R_H(T) 
> 0$ for hole--doped compounds and $R_H(T) < 0$ for electron--doped 
ones, although each of them has similar hole-like Fermi surface.
\item The $NMR$ relaxation rate, $1/T_1 \propto T^0$ 
for these materials. Remember that for a Fermi liquid, 
$1/T_1 \propto T$.
\item In conventional metals, one observes a Drude peak in the 
optical conductivity at $\omega = 0$, which broadens but persists 
for high temperatures. In contrast, in strongly correlated 
metals most of the spectral weight is in broad features at 
high frequencies. Furthermore, the Drude peak only exists 
at low temperatures.
\item In conventional metals, the thermopower is linear in 
temperature, has values less than $k_B/e \approx 87 \mu V/K$ 
and has the same sign as the charge carriers. In strongly 
correlated metals it can have a non--monotonic temperature 
dependence, can change sign and have values of the order of 
$k_B/e$.
\item The isotope exponent, $\alpha$', in $HTSC$ is unconventional 
in different respects. Optimally doped samples show a very small 
$\alpha'$ of the order of $0.05$ or even smaller, in contrast to 
the $BCS$ value of $1/2$. This unusually small value in connection 
with the high value of $T_c$ lead to early suggestions that the 
pairing interaction in the $HTSC$ cuprates might be predominantly 
electronic in origin with a possible small phononic contribution. 
However, to complicate things a little bit more, the isotope 
exponent factor, $\alpha'$, also shows an unusually strong 
doping dependence, reaching values of $1/2$, in some cases 
higher, in the underdoped, $T_c$ reduced, compounds.
\end{enumerate}

\indent Among one of the models to explain the {\it anomalous} 
properties of the $HTSC$ we mention the one of Varma et al\cite{Varma}, 
which tries to fit the linear resistivity data of the $HTSC$ by 
means of the quasi-particle lifetime of the form 
$\tau \approx 1/[T(\varepsilon_{k}-\mu)]$, where $T$ is the absolute 
temperature and $\mu$ is the chemical potential. This model 
is strictly phenomenological. There is a second model introduced 
by Anderson\cite{Phil}, whose starting hypothesis is the 
occurrence in two dimensional ($2-d$) systems of a state similar 
to the one from the one dimensional ($1 -d$) Luttinger 
liquid\cite{Mahan}. The point of view adopted by us is that the 
$HTSC$ are materials which can be treated by the Luttinger 
liquid model, specially in the underdoped regime. 
 Being so, we explore some consequences of this assumption.

\indent In such a non--Fermi system the one--particle spectral 
function $A(\varepsilon_{\vec{k}},\omega) = 
-1/\pi~Im[G(\varepsilon_{\vec{k}},\omega + i0^+)]$, 
where $G(\vec{k},\omega)$ is the one--particle Green function, 
satisfies the scaling property
\begin{equation}\label{scaling}
A(\Lambda\varepsilon_{\vec{k}},\Lambda\omega) = 
\Lambda^{\alpha-1} \times A(\varepsilon_{\vec{k}},\omega)~~~,
\end{equation}
\noindent where $\alpha$, the non--Fermi liquid parameter, 
is given by the condition $0~<~\alpha~<~1/2$. In the case of a Fermi 
liquid system $\alpha = 0$. We mention that Wen\cite{Wen}, as 
we do in the paper (Eq. (\ref{tie})), has shown that 
the exponent $\alpha$ is not 
universal, since it depends on the coupling constant between the 
electrons. In the case of the Hubbard model, $\alpha$ depends 
on the local Coulomb interaction. The model that we are 
going to study by means of the scaling relation 
(Eq. (\ref{scaling})) has poles and branch cuts. This scaling 
behaviour is a generalisation to dimensions $d > 1$ of the 
fermionic propagator from one dimension systems ($d = 1$). It 
has been showed by different authors\cite{4,5,6} that the 
one--particle Green function ($1PGF$) can be expressed as
\begin{equation}\label{gNFLGF}
G(\varepsilon_{\vec{k}},\omega) = 
\frac{g(\alpha)~e^{i\phi}~ \omega_c^{-\alpha}}
{(\omega-\varepsilon_{\sigma,\vec{k}})^{1/2}
(\omega - \varepsilon_{\rho,\vec{k}})^{1/2-\alpha}} ~~~,
~~~-\o_c<\o<\o_c
\end{equation}
\noindent where $\varepsilon_{\sigma(\rho),\vec{k}}$ represents 
the spin (charge) energy spectrum of the carriers, $\omega_c$ is 
frequency cutoff introduced to maintain the dimension of the 
$1PGF$ correct and  $g(\alpha)$, $e^{i\phi}$ are renormalization factors 
introduced in order to recover the usual properties of the Green's 
functions. Outside the interval $[-\o_c;\o_c]$ the Green function 
will have the usual $1/\o$ behaviour as $\o\ra\infty$.

\indent Eq. (\ref{gNFLGF}) was studied in some detail by 
Yin and Chakravarty\cite{YCh}. Let us mention that Eq. 
(\ref{scaling}) was used 
by Chakravarty and Anderson\cite{ChPhil} to obtain an interlayer 
tunneling Hamiltonian ($ILT$). Just recently, Chakravarty, Kee and 
Abrahams\cite{ChKAb} have used the interlayer tunneling 
Hamiltonian to explain the experiments of Basov et al\cite{Basov} 
on the $c$--axis optical sum rule in some $HTSC$. The structure of 
the paper is as follows.

\indent In Section \ref{II} we show that a band renormalization 
parameter $\gamma$ is necessary to satisfy the first order sum rule 
of Nolting\cite{Nolting}. The presence of $\gamma$ seems to us 
very natural since when correlations are present the free electronic 
band has to be renormalized too. In the same section, we find that 
$\alpha$, $\omega_c$ and $U$ are closely related. The parameter 
$\gamma$ is going to play an important role in the following 
development. In Section \ref{III} we have recalculated 
the following quantities: 1) the momentum distribution function, 
$n_c(\varepsilon_{\vec{k}})$ close to the effective chemical potential, 
namely, $\varepsilon_{\vec{k}} \approx \mu_{eff}$; 2) the 
superconducting critical temperature, $T_c$; and 3) the 
superconducting critical interaction strength, $\lambda_{cr}$. 
In Section \ref{IV} we present the calculation of the isotope 
exponent or coefficient, $\alpha'$, using the non--Fermi 
liquid one--particle Green function ($NFLGF$). 
In Section \ref{V} we present our 
conclusions and the outlook of our line of work. 

\indent While 
the semi phenomenological theory of Anderson\cite{Phil} 
suggests anomalous exponents, i.e., $\alpha \neq 0$, 
a {\it satisfactory} derivation of them and other 
details await further theoretical development\cite{genatios}. 
In Appendix \ref{a20} we have found a novel application of our 
theoretical treatment: two inequivalent coupled Hubbard layers 
($ICHL$). Among the new features found we mention the appearance 
of an energy gap in the off--diagonal one--particle Green 
function. We suggest that angle resolved photoemission spectroscopy 
($ARPES$) experiments should be set up to 
measure this gap foreseen for this type of materials ($Y_2Ba_4Cu_7O_{15} 
(247) \equiv YBa_2Cu_3O_{7} (123) + YBa_2Cu_4O_{8} (124)$). 
The results presented in this paper, specially the ones of 
Section \ref{II}, represent a first step along these lines, i.e., 
the use of the sum rules of Nolting\cite{Nolting} to determine 
the physical parameters of the theory. 

\section{Calculation of the parameters via sum rules}\label{II}
\indent The model we study is the Hubbard Hamiltonian\cite{HubbardI}
\begin{eqnarray}\label{Ham}
H = t_{i,j}
        c_{i\sigma}^{\dagger}c_{j\sigma}
   + \frac{U}{2} n_{i\sigma}n_{i\bar{\sigma}}   
   - \mu c^{\dagger}_{i\sigma}c_{i\sigma}~~,
\end{eqnarray}
where $c_{i\sigma}^{\dagger}$ ($c_{i\sigma}$) 
are creation (annihilation)  operators for particles  
with spin $\sigma$. $n_{i\sigma} \equiv 
c_{i\sigma}^{\dagger}c_{i\sigma}$, 
$U$ is the local interaction, $\mu$ the chemical
potential (we work in the grand canonical ensemble). We have 
adopted the Einstein convention for repeated 
indices, i.e., for the $N_s$ sites (labelled by $i$), the 
$z$ nearest-neighbour ($n.n.$) sites $j$ and for spin up and 
down ($\sigma = -\bar{\sigma} = \pm 1$). $t_{i,j} 
= -t$, for n.n. and zero otherwise.

\indent In this section we will use the first sum rules or 
moments of Nolting\cite{Nolting} applied to the 
Hubbard model of Eq. (\ref{Ham}) to find some conditions on the 
parameters of the theory. 
Before applying the sum rules, let us 
present the spectral function, $A(\vec{k},\omega)$. 
It is given by
\bea\label{A}
A(k,\o)&=&-\f{g(\a)}{\pi\o_c^\a}\left[
\f{\sin{[\phi-\pi(1-\a)]}}
{(\xi_{\vk}-\o)^{1/2-\a}(\eta\xi_{\vk}-\o)^{1/2}}
\Theta(\eta\xi_{\vk}-\o)\right.\nonumber\\
&+&\f{\sin{[\phi-\pi(1/2-\a)]}}
{(\xi_{\vk}-\o)^{1/2-\a}(\o-\eta\xi_{\vk})^{1/2}}
\Theta(\o-\eta\xi_{\vk})\Theta(\xi_{\vk}-\o)\nonumber\\
&+&\left.\f{\sin{\phi}}
{(\o-\xi_{\vk})^{1/2-\a}(\o-\eta\xi_{\vk})^{1/2}}
\Theta(\o-\xi_{\vk})\right]
\eea
if $0<\o<\o_c$. In this equation $\xi_{\vec{k}} \equiv 
\varepsilon_{\vec{k}} - \mu_{eff}$, $\eta=u_\rho/u_\sigma$ 
represents the ratio of the charge and spin velocities and $\Theta(x)$ 
is the usual theta function. The effective chemical potential is 
defined in Eq. (\ref{m1}). 
A similar equation can be obtain also for the case $-\o_c<\o<0$. 
The spectral function (Eq. (\ref{A})) has 
to satisfy the time reversal symmetry, 
condition which leads to $\phi=-\pi\a/2$, a result already obtained by 
Yin and Chakravarty\cite{YCh}. Eq. (\ref{A}) is assumed to be 
the many--body solution to the Hubbard Hamiltonian (Eq. (\ref{Ham})).

\indent The first moment (or zeroth order sum rule) 
$M_o(\vec{k})$, is given by
\begin{equation}\label{mo}
M_o(\vec{k})~ =~\int_{-\infty}^{+\infty}A(k,\omega)~d\omega~=~1~~~
\end{equation}
\indent Let us say that the condition given by Eq. (\ref{mo}) 
represents the equal--time anticommutation relation of fermions 
as pointed out by Yin and Chakravarty\cite{YCh}.  
Eq. (\ref{mo}) is valid for any fermionic theory, independent of the 
model used. We stress the fact that the fulfilment of Eq. (\ref{mo}) 
implies that our $1PGF$ describe fermion quasi--particles. 
Furthermore, the area of the distribution is one, 
i.e., it is normalised. Doing the integration of $A(k,\omega)$ 
and using the expression given in Eq. (\ref{A}), we find
\bea\label{intA}
\f{g(\a)}{\pi}\sin{\f{\pi\a}{2}}
\left\{\f{1}{\a}\left[\left(1+\f{\eta\xi_{\vk}}{\o_c}\right)^\a+
\left(1-\f{\eta\xi_{\vk}}{\o_c}\right)^\a \right]+
(1-\eta)\f{\xi_{\vk}}{\o_c}\f{\a-1/2}{\a-1}
\left[\left(1+\f{\eta\xi_{\vk}}{\o_c}\right)^{\a-1}-
\left(1-\f{\eta\xi_{\vk}}{\o_c}\right)^{\a-1}\right]\right\}\nonumber\\
+\f{g(\a)}{\pi}(1-\eta)^\a\f{\xi_{\vk}^\a}{\o_c^\a}
\left[\cos{\f{\pi\a}{2}}B(1/2,\a+1/2)+\sin{\f{\pi\a}{2}
\f{\a^2-3/2\a+1}{\a(\a-1)}}\right]=1
\eea
where $B(x,y)$ is the usual beta Euler function. If we restrict 
ourselves to regions close to the effective chemical 
potential, i.e., $\xi_{\vec{k}} \approx 0$, then the normalisation 
factor, $g(\alpha)$, is independent of $\vec{k}$ and is given by
\begin{equation}\label{g(a))}
g(\alpha) \approx \frac{\pi \alpha}{2\sin(\frac{\pi\alpha}
{2})}~~~,
\end{equation}
which reduces to 1 when $\alpha \rightarrow 0$. 

The first order sum rule is given by
\begin{equation}\label{m1}
M_1(\vec{k})~ =~\int_{-\infty}^{+\infty}~\omega A(k,\omega)~d\omega~=~
\varepsilon_{\vec{k}}-\mu + \rho U \equiv \xi_{\vec{k}}~~~;~~~
\mu_{eff} \equiv \mu - \rho U~~~.
\end{equation}
\indent The integral equals to $\varepsilon_{\vec{k}}-\mu + \rho U$ 
is an exact result within the working scheme of the Hubbard model 
(model dependent\cite{Wen}). In Eq. (\ref{m1}), $\rho$ is the 
carrier number per lattice site and per spin. We work in the 
paramagnetic phase, namely, $\rho_{\uparrow} = \rho_{\downarrow} 
= \rho$. Combining Eqs. (\ref{A},\ref{m1}), i.e., doing the 
integral of $\omega~A(k,\omega)$ to find the centre of the 
distribution, we get 

\begin{eqnarray}\label{intwA}
\f{g(\a)}{\pi\o_c^\a}\sin{\f{\pi\a}{2}}&\times&
\left\{-\f{\o_c^{\a+1}}{\a+1}
\left[\left(1+\f{\eta\xi_{\vk}}{\o_c}\right)^{\a+1}-
\left(1-\f{\eta\xi_{\vk}}{\o_c}\right)^{\a+1}\right]\right.\nonumber\\
&-&\xi_{\vk}
[(1-\eta)(\a-1/2)-\eta]\f{\o_c^\a}{\a}
\left[\left(1+\f{\eta\xi_{\vk}}{\o_c}\right)^\a+
\left(1-\f{\eta\xi_{\vk}}{\o_c}\right)^\a\right]\nonumber\\
&+&\left.\xi_{\vk}^2\eta(1-\eta)(\a-1/2)\f{\o_c^{\a-1}}{\a-1}
\left[\left(1+\f{\eta\xi_{\vk}}{\o_c}\right)^{\a-1}-
\left(1-\f{\eta\xi_{\vk}}{\o_c}\right)^{\a-1}\right]\right\}\nonumber\\
+\f{g(\a)}{\pi\o_c}\xi_{\vk}^{\a+1}(1-\eta)^\a&\times&
\left\{\cos{\f{\pi\a}{2}}
\eta 
B\left(\a+\f{1}{2},\f{1}{2}\right)
F\left(-1,\f{1}{2};\a+1;-\f{1-\eta}{\eta}\right)
\right.\nonumber\\
&+&\left.\sin{\f{\pi\a}{2}}
\left[\f{\eta(\a-1/2)}{\a-1}-\f{1-\eta}
{\a+1}-\f{\eta-(\a-1/2)(1-\eta)}{\a}\right]\right\}=\xi_{\vk}
\end{eqnarray}
\noindent where $F(\alpha,\beta;\gamma;z)$ is the hypergeometric 
function. Using an expansion around $\xi_{\vec{k}} \approx 0$ as 
previously, we find that 
\be\label{imposibil}
\f{2g(\a)\sin{[\pi\a/2]}}{\pi\a}
\left[\eta-(1-\eta)\left(\a-\f{1}{2}\right)-\eta\a\right]\xi_{\vk}
\approx \xi_{\vk}
\ee
which leads to the following relation between the anomalous 
coefficient $\a$ and the spin-charge characteristic ratio $\eta$
\be
\label{wrong}
\a=\f{\eta-1}{2}
\ee
If we analyse Eq. (\ref{wrong}) we can see that as long as 
$0<\a<1/2$ and $0<\eta<1$ the only possibility is that $\a=0$ and 
$\eta=1$, which is actually characteristic for the usual non 
interacting Fermi 
liquid. As a conclusion, in order to satisfy the first order sum rule 
we have to introduce a new coefficient $\gamma$ related to the 
band renormalization factor. With this coefficient the energy $\xi_{\vk}$ 
will be renormalized as $\gamma\xi_{\vk}$. By applying again the 
first order sum rule, we get
\be\label{gama}
\gamma=\f{2}{\eta+1-2\a}
\ee
so the band renormalization factor will be a function of the two previous 
parameters $\a$ and $\eta$. We have to mention that the introduction 
of the band renormalization factor does not affect the time reversal symmetry 
and the zeroth order sum rule. We remark 
that this relation had not been previously calculated in the 
literature because only the first sum rule (Eq. (\ref{mo})) 
had been used. Now, we are going further and we will apply 
the second order sum rule or the width of the distribution 
function, $A(\vec{k},\omega)$, to calculate 
$\alpha$. Indeed, what we will find is that this 
sum rule imposes a condition on the three remaining parameters 
of the theory, namely, $\omega_c$, $\alpha$ and $U$. The 
interaction $U$ is coming from the Hubbard model. As the parameter 
$\alpha$ is not really independent of the Coulomb interaction,  
our result evidently departs from the approach adopted by 
\c{T}ifrea\cite{Tifrea1} in his Ph. D. thesis and related 
works\cite{Tifrea2} (see the discussion after Eq. (\ref{tie})).

\indent The second order moment, $M_2(\vec{k})$, is given 
as 
\begin{equation}\label{m2}
M_2(\vec{k})~ =~\int_{-\infty}^{+\infty}~\omega^2 A(k,\omega)~d\omega~=~
(\varepsilon_{\vec{k}}-\mu)^2 + 2 U (\varepsilon_{\vec{k}}-\mu) + 
\rho U^2 \equiv \xi_{\vec{k}}^2 + \rho(1-\rho)U^2~~~.
\end{equation}
\indent This is an exact relation and the model dependence 
is visible thru the local Coulomb interaction. Performing the 
integral of $\omega^2 A(k,\omega)$ to find the width 
of the distribution and using the definition 
of the one--particle spectral function given in Eq. (\ref{A}), we find that 
in the limit $\xi_{\vk}\approx 0$
\be
\f{2g(\a)\sin{[\pi\a/2]}}{\pi\a}\f{\a}{\a+2}\o_c^2\approx \rho (1-\rho) U^2
\ee
which gives
\begin{equation}\label{tie}
\frac{\alpha}{\alpha+2} \approx \rho (1-\rho) 
\left(\frac{U}{\omega_c}\right)^2~~~.
\end{equation}
\indent Thus, as we already pointed out, Eq. (\ref{tie}) puts a 
strong condition on the remaining physical variables of the 
theory. We immediately see that $\alpha = 0$ for $U \equiv 0$, 
as it should be. Also, another parameter entering in the 
constraint is the electron number/spin. As $0 < \alpha < 1/2$, 
we see that $U_{max} \approx \omega_c$, for $\rho \neq 0;~1$. 
Here we appreciate the difference with the work of 
\c{T}ifrea\cite{Tifrea1} in his Ph.\ D.\ thesis where he 
calculates the superconducting critical temperature, $T_c$, 
in the presence of repulsive local Coulomb interaction. The 
non--Fermi liquid parameter, $\alpha$, continues to be 
a free parameter in \c{T}ifrea's approach. Before we leave this 
Section, we say that our energy scales are ordered in the following 
way, $T_c < \Delta(0)  < \omega_D << \omega_c << D$, where 
$\Delta(0)$ is the superconducting gap at zero temperature, 
$\omega_D$ is the Debye frequency giving origin to the 
superconducting critical temperature, $T_c$. 

\section{Dynamical and global quantities}\label{III}
\indent In this Section we will calculate some {\it dynamical}  
properties of the theory, namely, the momentum distribution function, 
$n_c(\vec{k})$, in the normal phase at $T = 0$. Also, we 
will recalculate the superconducting critical temperature, 
$T_c$, and the superconducting critical interaction strength, 
$\lambda_{cr}$, paying duly attention to the presence of the 
new parameter of the theory, $\gamma = 2/(\eta+1-2\alpha)$ 
(Eq. (\ref{gama})).

\subsection{Calculation of $n_c(\varepsilon_{\vec{k}})$ at $T = 0$}
\label{III.2}
\indent The correlated momentum distribution function is 
given by the following expression
\begin{equation}\label{nk}
n_c(\varepsilon_{\vec{k}}) \equiv \int_{-\infty}^{+\infty}
~d\omega~\frac{A(\varepsilon_{\vec{k}},\omega)}
{\exp(\omega/T) + 1} 
\end{equation}

\indent At $T = 0$ we have to look carefully to this integral 
because as long as $\o>0$ the exponential function is infinity, 
which implies a zero contribution from the integral. We still have 
to integrate over the region $\o<0$ where the exponential 
function is zero. 

\indent Performing the integrals we end up with
\begin{eqnarray}\label{nkfinal}
\lim_{|\xi_{\vk}| \rightarrow 0} 
n_c(\varepsilon_{\vec{k}}) = \frac{1}{2}\left\{
1-sign(\xi_{\vk})
\left[\left(\f{2\eta}{\eta+1-2\a}\f{|\xi_{\vk}|}{\o_c}\right)^\a
\right.\right.&-&2^{1-\a}\a(\a-1/2)B\left(\f{1}{2},1-\a\right)
\left(\f{2(1-\eta)}{\eta+1-2\a}\f{|\xi_{\vk}|}{\o_c}\right)^\a
\nonumber\\
&+&\left.\left.\a(\a-1/2)\f{1-\eta}{\eta}\left(\f{2\eta}{\eta+1-2\a}
\f{|\xi_{\vk}|}{\o_c}\right)f(\a,\eta)\right]\right\}
\end{eqnarray}
where $f(\a,\eta)=\int_0^1dz z^{-1/2}[z+(1-\eta)/\eta]^{\a-3/2}$.

\indent As we see from Eq. (\ref{nkfinal}), the correlated 
momentum distribution function has been calculated close to 
the effective chemical potential. Also, we observe that for 
$\alpha = 0$ and $\eta=1$ we recover the jump at the 
chemical potential, 
as it is the case for a Fermi liquid. For $\alpha \neq 0$ 
and $\eta=1$, 
this jump has gone away, but the derivative at the effective chemical 
potential is discontinuous. A calculation of the renormalization factor 
$Z$, defined as $Z=n_c(\xi_{\vk}^+)-n_c(\xi_{\vk}^-)$, gives $Z=0$, 
a result which 
implies that our theory is a non-Fermi liquid one. In the other 
case, $\a=0$ and $\eta\neq 1$, we obtain the case of a Fermi 
liquid with spin-charge separation ($Z=1$). 
Another conclusion that we reach 
by looking at Eq. (\ref{nkfinal}) is that the parameter 
$\gamma= 2/(\eta+1-2\alpha)$ modifies the results of Yin and 
Chakravarty\cite{YCh} in the following way: in order 
to study non-Fermi liquid systems we have to consider that 
the frequency cutoff is effectively smaller, i.e., 
$\omega_c \rightarrow (1-\alpha)\omega_c$. 

\subsection{Calculation of $T_c$ and $\lambda_c$ in an 
s--wave superconductor}\label{III.3}

In the following we will restrict ourself to the study of the 
non-Fermi liquid system ($\a\neq 0$, $\eta=1$) where the Green's 
function according to our previous calculations is given by
\be
G_0(k,\o)=\f{g(\a)e^{-i\pi\a/2}}
{\o_c^\a(\o-\gamma\xi_{\vk})^{1-\a}}
\label{GF}
\ee
with $\gamma=1/(1-\a)$. The order parameter equation in the 
framework of the Gorkov equations is given by:
\be
\label{gap}
1=V\sum_{\vk}\f{1}{\b}\sum_{\o_n}
\f{1}{G_0^{-1}(k,i\o_n)G_0^{-1}(-k,-i\o_n)-|\D_{\vk}|^2}
\ee
where $\b=1/T$. The critical temperature will be obtained from 
Eq. (\ref{gap}) with the condition $\D_{\vk}\ra 0$, and the 
difficult problem of evaluating the sum over the Matsubara 
frequency will be solve by using a contour integral similar 
with the one used in Ref. \onlinecite{Tifrea1,Tifrea2}. In the 
limit $\b\o_D\ll 1$ the critical temperature can be obtained exactly 
as
\be
\label{Tc}
T_c^{2\a}=\f{1}{C(\a)}\left[D(\a)\left(\gamma\o_D\right)^{2\a}
-\f{\gamma}{g^2(\a)}\f{\o_c^{2\a}}{\l A(\a)}\right]
\ee
where $\l=1/N(0)V$ and $A(\a)$, $C(\a)$ and $D(\a)$ has the same 
meaning as in the paper 
of Muthukumar et al. \cite{Tifrea2}. We have to mention that our critical 
temperature is different from the one obtained in Ref. \onlinecite{Tifrea2}, 
and include the renormalization factor $g(\a)$ and the band 
renormalization factor. As a result we can see a decrease of the critical 
temperature due to the effective Debye frequency $\o_D^{eff}=\gamma\o_D$. 
We also obtained a modified critical coupling constant $\l_{cr}$
\be
\label{lc}
\l_{cr}=\f{\gamma}{g^2(\a)}\left(\f{\o_c}{\gamma\o_D}\right)^{2\a}
\f{1}{A(\a)D(\a)}
\ee
which by the same reason of an enhanced effective Debye frequency 
seems to be enhanced. In Fig.1 we present a plot of the critical 
coupling constant versus the non-Fermi parameter $\a$. As we can see 
in the limit $\a\ra 0$ we recover the usual BCS result\cite{Tifrea1} 
($\l_{cr}=0$).

\section{The isotope effect for non--Fermi liquids}\label{IV}
\indent The isotope effect exponent is an important physical 
parameter since, in 
the low temperature superconductors ($LTSC$), it was used to determine the 
origin of the pairing mechanism. In the $HTSC$ the isotope 
effect has been widely discussed, in particular to detect the 
influence of the phonon degrees of freedom on $T_c$. 
As Kishore\cite{Ramisotope} 
points out, the isotope effect in the cuprates is sensible to 
various factors like: 1) the form of the free density of states, 
2) the Coulomb interaction, 3) the carrier concentration, 
4) presence of impurities, 5) anharmonicity and 6) the symmetry 
of the order parameter, among others. From our Eq. (\ref{tie}) 
we can account for the first three dependences. Here we will calculate 
$T_c$ and the isotope exponent, $\alpha'$, for a $s$--wave 
superconductor. Thus, our expression for $\alpha'$ could be 
approximately applicable to the cuprates to fit the data. 
We believe that the numerical value of global 
quantities, like $\alpha'$, do not depend too much on the 
symmetry of the order parameter. It appears that $La_{2-x}SrCuO_4$ 
presents a $s$--wave symmetry order parameter. So, it is the 
natural candidate to apply the ideas worked out here. On the 
other side, $Bi_2Sr_2CaCu_2O_8$\cite{Fermisurface} holds a 
$d$--wave symmetry order parameter.

\indent The isotope effect exponent is defined as 
$T_c \propto M^{-\alpha'}$. Then, it is given by
\begin{equation}\label{alpha'}
\alpha' = - \frac{\partial \ln T_c}{\partial \ln M} ~~~,
\end{equation}
\noindent where $M$ is the isotope mass. Another relation 
that we will use is the relation between the Debye frequency 
and the isotope mass, namely, $\omega_D \propto M^{-1/2}$. 
Now, Eq. (\ref{Tc}) can be re-written as   

\begin{eqnarray}\label{1/lambda}
\frac{1}{\lambda}&=&g^2(\a)A(\a) F(\o_D,\a,T_c)\nonumber\\
F(\o_D,\a,T_c)&=&\left[\gamma^{2\a-1}D(\a)\left(\f{\o_D}{\o_c}\right)^{2\a}
-\f{C(\a)}{\gamma}\left(\f{T_c}{\o_c}\right)^{2\a}\right]
\end{eqnarray}
\indent Deriving Eq. (\ref{1/lambda}) with respect to $M$ we
get that
\begin{equation}\label{formalalpha'}
\alpha' = - \frac{\omega_D~\left(\frac{\partial F}{\partial \omega_D}
\right)}
{2 T_c~\left(\frac{\partial F}{\partial T_c}\right)}~~~.
\end{equation}
\indent Let us mention that Eq. (\ref{formalalpha'}) is valid 
for the case that $\alpha \neq f(\omega_D)$. The case  
$\alpha = f(\omega_D)$ will be discussed later on (Eq. 
(\ref{generalalpha'})). Performing 
the partial derivatives, we come down to the following 
expression
\begin{equation}\label{finalalpha'}
\alpha'=\frac{1}{2}\left[1-\f{\gamma}{g^2(\a)}
\left(\f{\o_c}{\gamma\o_D}\right)^{2\a}
\f{1}{\l A(\a)D(\a)}\right]^{-1} 
\end{equation}
\indent From Eq. (\ref{finalalpha'}) we gain the $BCS$ case, 
i.e., $\alpha' = 1/2$ when $\alpha = 0$. 
In Fig.2 we plot the isotope coefficient versus the non-Fermi 
parameter $\a$ for different values of the coupling constant. 
As we expected there is deviation for the usual BCS result as 
$\a$ increases. Also we mention that the isotope coefficient 
depends on the coupling constant, a result similar with the one 
obtained for the $2\D(0)/T_c$ ratio \cite{Tifrea1}.

\indent While leaving this Section, let us mention that if 
$\alpha$ would depend on $\omega_D$, then the isotope 
effect exponent should be obtained from the following 
expression
\begin{equation}\label{generalalpha'}
\alpha' = - \frac{\omega_D~
\left[\frac{\partial F}{\partial \omega_D} + 
\frac{\partial F}{\partial \alpha}~
\frac{\partial \alpha}{\partial \omega_D} \right]}
{2 T_c~\left(\frac{\partial F}{\partial T_c}\right)}~~~.
\end{equation}
\indent We could make $\alpha = f(\omega_D)$ by 
choosing $\omega_c \propto \omega_D$, as \c{T}ifrea 
has done in his Ph.\ D. thesis\cite{Tifrea1}. This 
contribution is important to get full agreement with the 
experimental data, namely, that $\alpha' \approx 0$ at 
optimal doping. We argue that the lowering of $\alpha'$ from 
the $BCS$ result should come from the second contribution 
in the numerator in Eq. (\ref{generalalpha'}). 
We leave this task for the future. However, the result found 
in the present section should be applicable in the underdoped 
regime of the $HTSC$'s. 

\section{Conclusions and outlook}\label{V}
\indent We have applied the first three sum rules of 
Nolting\cite{Nolting} for the $NFLGF$ of Eq. (\ref{A}), 
which is assumed to be a solution of the Hubbard model 
(Eq. (\ref{Ham})) for frequencies $|\omega| \leq \omega_c$. 
This $NFLGF$ is anomalous due to the presence of the 
non--Fermi liquid exponent, $\alpha$. Due to the 
requirement that the first order sum rule of the spectral 
function be satisfied, a new parameter has to be called 
for. $\gamma$, the so called mass renormalization factor 
plays an important role in the theory. Due to 
its presence we have recalculated some results found in the 
literature and pointed out the role of $\gamma$. Also, a 
new quantity, the isotope effect exponent or $\alpha'$, 
has been calculated for the first time for 
non--Fermi liquid systems. $\alpha'$ reduces to the $BCS$ case 
when the Coulomb interaction is zero, namely, for $\alpha 
= 0$. The fact that $\alpha \propto U^2$ comes out from the 
application of the second order sum rule to the spectral 
function. Due to Eq. (\ref{tie}), 
$\alpha$ is a model dependent parameter, a result 
found previously in the literature by Wen\cite{Wen}. 
According to our view, $T_c$ decreases with $\alpha$\cite{Tifrea1} 
and $\alpha \propto U^2$, then we can conclude that the 
local Coulomb repulsive interaction is detrimental to 
superconductivity. This result has been found in other 
approaches\cite{last1,last2}.

\indent We would like to point out that our spectral function 
(Eq. (\ref{A})) does not reduce to the well known Dirac delta 
function when $\alpha = 0$, $\eta=1$. 
In order to re-obtain the usual delta 
form of the spectral function we have to set $\a=0$ and $\eta=1$ in 
our starting Green function. Even without doing that for $\a=0$ and 
$\eta=1$ the spectral function (Eq. \ref{A}) will satisfy the usual 
scaling relation for a Fermi liquid system $A(\L[k-k_F],\L\o)=
\L^{-1}A([k-k_F],\o)$. We used this property in some 
parts of the paper and we have recovered known features, like: 
1) the jump at $\mu$ of the momentum distribution function, 
$n_c(\vec{k})$, which is a Fermi liquid behaviour;  
2) the superconducting critical interaction is zero, 
namely, $\lambda_{cr} = 0$, and 3) the 
$BCS$ isotope exponent, $\alpha' = 1/2$. 

\indent In Appendix \ref{a20} we have applied the 
formalism developed in this 
paper to two inequivalent coupled Hubbard layers ($ICHL$) 
extending the calculation of Yin and Chakravarty\cite{YCh} 
for equivalent planes. Hildebrand et al\cite{Hanke} 
have applied the FLEX formalism to $ICHL$'s. In particular, 
our expressions for $G_{11}(\vec{k},i\omega_n)$, 
$G_{22}(\vec{k},i\omega_n)$ and $G_{12}(\vec{k},i\omega_n) = 
G_{21}(\vec{k},i\omega_n)$ are different from the ones of 
Ref.\cite{YCh} due to the presence of a shift in the 
effective chemical potential, i.e., $\mu_{1,eff} = \mu_{2,eff} + \delta$.  
Experiments should be designed to detect the results found 
in our work (See Appendix \ref{a20}). In particular, the 
presence of the theoretical gap in off--diagonal one--particle 
spectral function, $A_{12}(\vec{k},\omega)$, calls for 
$ARPES$ experiments to be performed in these 
materials\cite{Shen}. 

\indent We have assumed 
that the non-Fermi liquid parameter, $\alpha$, is independent 
of $\vec{k}$. This seems not to be the case as Meden\cite{Meden} 
has pointed out. He concludes that the asymptotic behaviour of the 
one--particle Green functions of Luttinger liquids at large 
space--time distances is not {\it universal}. Namely, along 
certain directions the exponent of the asymptotic power 
law is not given by the Luttinger liquid parameters. Due to 
this consideration, $\alpha$ could depend on $\vec{k}$. This 
possibility is outside the scope of the present paper. Among 
one of the possibilities we would like to explore is the 
calculation of pressure effects\cite{Evandro} in some 
$HTSC$ materials, when correlation is important, i.e., 
$\alpha \neq 0$. Another possibility worth considering, for the 
case of inequivalent coupled Hubbard layers, is $\alpha_1 
\neq \alpha_2$, as the two non--Fermi liquid parameters on the 
two planes. This approach is likely more demanding than the 
one followed in the paper. We leave for the future the 
self-consistent calculation of $\rho_1$ and $\rho_2$ on the 
two layers (See Eq. (\ref{123})). On the other hand, the presence 
of the mass renormalization parameter, $\gamma$, will modify 
the zero temperature order parameter, $\Delta(0)$. However, 
its calculation has been left out of the present work. 
Of course, with $\Delta(0)$, we could calculate the 
ratio $2\Delta(0)/T_c = f(U) \neq 3.5$, seeing its dependence on $U$ (or 
$\alpha$).  Another aspect which should be addressed 
in the future is the superconducting properties of two 
inequivalent coupled Hubbard layers. In this case, our mean 
field Hamiltonian becomes a $4\times4$ matrix in the 
Nambu formalism, giving rise to pairing on the planes and 
perpendicular to them. We would like to end by saying that the 
nature of the superconducting transition is strongly 
related to how anomalous (non--Fermi liquid like) the normal 
state spectral function is, and as such, is dependent 
upon the doping level\cite{condensation}. The anomalous 
properties of the normal state spectral function are 
visible in the underdoped regime of the $HTSC$.\\

\noindent {\bf Acknowledgements}\\
\indent We thank Prof.\ R.\ Kishore, Prof.\ E.\ V.\ L.\ de Mello, Prof.\ 
A.\ A.\ Schmidt, Prof. H.\ Beck, Prof.\ L.\ D.\ Almeida, 
Prof.\ F.\ Kokubun, Prof.\ M.\ Cri\c{s}an and Dr. I.\ 
C.\ Ventura for interesting discussions. Two of the authors (JJRN and SGM) 
thank partial support from FAPERGS--Brasil (Project 98/0701.1), 
CONICIT--Venezuela (Project F-139) and CNPq--Brasil. One of us (IT) 
gratefully acknowledge financial support from INFM Italy under the project 
PRA-HTSC (1999). We thank Prof.\ M.\ D.\ 
Garc\'{\i}a Gonz\'alez for reading the manuscript.

\appendix

\section{Non--Fermi liquid one--particle Green functions for 
two inequivalent coupled Hubbard layers}\label{a20}
\indent In Ref.\cite{Hanke}, Hildebrand et al have studied the 
case of two inequivalent coupled Hubbard layers for the case 
of $Y_2Ba_4Cu_7O_{15} (247) \equiv YBa_2Cu_3O_{7} (123) + 
YBa_2Cu_4O_{8} (124)$. We will re--study this system using 
$NFLGF$ for each one of the layers with the purpose of finding 
new theoretical consequences of this assumption. For example, 
we are going to compare with the results of Refs.\cite{YCh,Hanke}. 
The Hamiltonian of the system is given by
\[
H = 
\left(
\begin{array}{rr}
G_1^{-1}(\varepsilon_{\vec{k}},i\omega_n) & ~~t_{\bot}~~~~~ \\
~~t_{\bot}~~~~~ & G_2^{-1}(\varepsilon_{\vec{k}},i\omega_n) 
\end{array} \right)~~~,
\]
\noindent where $t_{\bot}$ is the coupling matrix element 
between the two inequivalent Hubbard layers and 
$G_j(\varepsilon_{\vec{k}},i\omega_n)$, $j =1,2$, are the 
normal state non--Fermi liquid one-particle Green functions of 
the layers.  They are given by

\begin{eqnarray}\label{twoplanes}
G_j(\varepsilon_{\vec{k}},i\omega_n) = \frac{g(\alpha)}{
\omega_c^{\alpha} e^{\pm i\pi\alpha/2}}~\frac{1}{\left(i\omega_n-\gamma 
\eta_{j,\vec{k}}\right)^{1-\alpha}}~~~\Theta(\pm \omega_n)~~~,
\end{eqnarray}
\noindent with $\eta_{1,\vec{k}} \equiv \varepsilon_{\vec{k}} - 
\mu_{eff}$ and $\eta_{2,\vec{k}} \equiv \eta_{1,\vec{k}} - 
\delta$, i.e., we have included a shift between the two effective chemical 
potentials. A simple calculation allows us to calculate 
the diagonal and off-diagonal one--particle Green functions. They 
are 
\begin{eqnarray}\label{G11-G22-G12}
G_{jj}(\varepsilon_{\vec{k}},i\omega_n) = 
\frac{G^{-1}_{\bar{j}}(\varepsilon_{\vec{k}},i\omega_n)}{
\left(G^{-1}_1(\varepsilon_{\vec{k}},i\omega_n) 
G^{-1}_2(\varepsilon_{\vec{k}},i\omega_n) 
-t^2_{\bot}\right)}~~~&;&~~~j = 1,2~~;~~\bar{1} = 2~;~\bar{2} = 1
\nonumber \\
~~~\nonumber \\
G_{12}(\varepsilon_{\vec{k}},i\omega_n) = 
\frac{t_{\bot}}{
\left(G^{-1}_1(\varepsilon_{\vec{k}},i\omega_n) 
G^{-1}_2(\varepsilon_{\vec{k}},i\omega_n) 
-t^2_{\bot}\right)}~~~&,&~~~G_{21}(\varepsilon_{\vec{k}},i\omega_n) = 
G_{12}(\varepsilon_{\vec{k}},i\omega_n)
\end{eqnarray}
\indent The excitation spectrum is determinated by the roots of 
the denominator of Eq. (\ref{G11-G22-G12}). There are branch 
points at $\omega = \gamma \eta_{j,\vec{k}}$. Let us assume that 
$\eta_{1,\vec{k}} > \eta_{2,\vec{k}} > 0$. Thus, for $\omega > 0$, 
we must divide the 
complex plane in six regions named $I$--$II$--$III$--$IV$--$V$--$VI$, as 
shown in Fig. 3.a. The poles are given by the solutions of
\begin{equation}
e^{i\pi\alpha} \omega_c^{2\alpha} g^{-2}(\alpha)\left(\omega - 
\gamma \eta_{1,\vec{k}} + i0^+\right)^{1-\alpha}\left(\omega - 
\beta \eta_{2,\vec{k}} + i0^+\right)^{1-\alpha} = 
t_{\bot}^2 ~e^{2i~n~\pi} ~~~;~~~n~~an~~integer~~~,
\end{equation}
Following the analysis performed by Yin and Chakravarty\cite{YCh} 
we conclude that, for $\eta_{1,\vec{k}} > \eta_{2,\vec{k}} > 0$, there 
are solutions only in the regions denoted by $I$ and $V$. 
In the case that $0 < \eta_{1,\vec{k}} > 
\eta_{2,\vec{k}} < 0$, still for $\omega > 0$, we divide the complex 
plane in four regions $I$--$II$--$III$--$IV$ as shown in Fig. 3.b. 
Similarly to the previous analysis, we conclude that there is 
solution in region $I$ only. The analysis
is similar for the case $\eta_{1,\vec{k}} < \eta_{2,\vec{k}} < 0$. 

\indent The poles are localized at
\begin{equation}\label{poles}
\omega_{1,2} = \frac{\gamma(\eta_{1,\vec{k}} + \eta_{2,\vec{k}}) 
\pm \left[\gamma^2\left(\eta_{1,\vec{k}} - \eta_{2,\vec{k}}\right)^2 +
4t^2_{\bot,eff}~e^{\pm \left(\frac{i\pi\alpha}{(1-\alpha)}
\right)}\right]^{1/2}}{2}~~~;~~~t_{\bot,eff} \equiv g(\alpha)~t_{\bot}
~\left(\frac{g(\alpha)t_{\bot}}{\omega_c}\right)^{\frac{\alpha}{
(1-\alpha)}}~~~.
\end{equation}
\noindent From Eq. (\ref{poles}) we recover the case of equivalent 
coupled planes of Yin and Chakravarty\cite{YCh} by making 
$\eta_{1,\vec{k}} = \eta_{2,\vec{k}}$. Now, we are in a position 
of calculating the one--particle spectral functions, namely, 
$A_{i,j}(\vec{k},\omega)$, with $i,j = 1,2$. The results are
(with $\gamma\eta_{1,\vec{k}} > \gamma\eta_{2,\vec{k}}$)
\[
A_{11}(\vec{k},\omega) = 
\frac{\sin(\pi\alpha/2)~X_2^{1-\alpha}}{\pi t_{\bot}}\left\{
\begin{array}{rl}
\frac{\left(1 + \left(X_2~X_1\right)^{1-\alpha}\right)}{\left(1 + 
\left(X_2~X_1\right)^{2(1-\alpha)}
- 2\cos\left(\pi\alpha\right)\left(X_2~X_1\right)^{1-\alpha}\right)} & 
\mbox{~~if ~~~$\omega 
< \gamma \eta_{2,\vec{k}} < \gamma \eta_{1,\vec{k}}$ or 
$\gamma \eta_{2,\vec{k}} < \gamma \eta_{1,\vec{k}} < \omega$}; \\
~~~~\\
\frac{1}{\left(1 +  \left(X_2~X_1\right)^{1-\alpha}\right)}~~~~~ & 
\mbox{~~if 
~~~$\gamma \eta_{2,\vec{k}} < \omega < 
\gamma \eta_{1,\vec{k}}$}
\end{array}
\right.
\]
\noindent where $X_j \equiv \frac{|\omega - \gamma \eta_{j,\vec{k}}|}
{|t_{\bot,eff}|}$. 
\[
A_{22}(\vec{k},\omega) = 
\frac{\sin(\pi\alpha/2)~X_1^{1-\alpha}}{\pi t_{\bot}}\left\{
\begin{array}{rl}
\frac{\left(1 + \left(X_2~X_1\right)^{1-\alpha}\right)}
{\left(1 + \left(X_2~X_1\right)^{2(1-\alpha)}
- 2\cos\left(\pi\alpha\right)
~\left(X_2~X_1\right)^{1-\alpha}\right)} & \mbox{~~if ~~~$\omega 
< \gamma \eta_{2,\vec{k}} < \gamma \eta_{1,\vec{k}}$ or 
$\gamma \eta_{2,\vec{k}} < \gamma \eta_{1,\vec{k}} < \omega$}; \\
~~~\\
\frac{1}{\left(1 +  \left(X_2~X_1\right)^{1-\alpha}\right)}~~~~~~ & 
\mbox{~~if ~~~$\gamma 
\eta_{2,\vec{k}} < \omega < \gamma \eta_{1,\vec{k}}$}
\end{array}
\right.
\]

\[
A_{12}(\vec{k},\omega) = A_{21}(\vec{k},\omega) = 
\frac{\sin(\pi\alpha)~\left(X_1~X_2\right)^{1-\alpha}}
{\pi t_{\bot}}\left\{
\begin{array}{rl}
\frac{1}
{\left(1 + \left(X_2~X_1\right)^{2(1-\alpha)}
- 2\cos\left(\pi\alpha\right)
~\left(X_2~X_1\right)^{1-\alpha}\right)} & 
\mbox{~~if 
~~~$\omega > \gamma \eta_1$}; \\
~~\\
\frac{-1}
{\left(1 + \left(X_2~X_1\right)^{2(1-\alpha)}
- 2\cos\left(\pi\alpha\right)
~\left(X_2~X_1\right)^{1-\alpha}\right)} & 
\mbox{~~if 
~~~$\omega < \gamma \eta_2$}; \\
~~\\
0 ~~~~~~~ & \mbox{~~if ~~~$\gamma \eta_1 < \omega < \gamma \eta_2$}
\end{array}
\right.
\]

\indent In Figures 4 we show the behaviour of the diagonal and 
off--diagonal one-particle spectral functions. 
From Figure 4.a we observe that the symmetry of the 
diagonal spectral function, $A_{11}(\vec{k},\omega)$, is 
lost around the frequency $\omega = \gamma \eta_{2,\vec{k}} = 
0.9$. To realize the new feature, we refer the reader to 
the respective figure ($F_{11}(x)$) in Ref.\cite{YCh}. 
Also, the symmetry of $F_{22}(x)$ in Ref.\cite{YCh} is lost 
around $\omega = \gamma \eta_{1,\vec{k}} = 1.2$. In consequence, 
$ARPES$ experiments shining light on layer 1 or layer 2 should 
detect these fine details. Namely, we will have two 
symmetry breaking,  $1 \rightarrow 1$ and  
$2 \rightarrow 2$. Now, from Figure 4.b we see that 
there is an interval of energy where $A_{12}(\vec{k},\omega) 
= 0$. This feature was not obtained for the 
off--diagonal one--particle spectral function for 
equivalent planes (See $F_{12}(x)$ in 
Ref.\cite{YCh}). Another aspect which we would like to 
point out is the fact that in the one--particle spectral 
functions there is not an unique variable to describe the 
{\it data}. We have two relevant energy scales in the 
problem, for each value of $\vec{k}$. 
For the type of materials we are studying it seems 
natural to find, on theoretical grounds, a gap in the energy 
spectrum. $ARPES$ experiments 
should be designed to detect the gap found from our 
expressions\cite{Shen}. 
We mention that the gap found in $A_{12}(\vec{k},\omega)$ depends 
on the relative carrier number in the two inequivalent coupled 
Hubbard layers. Therefore, the presence of a shift between the 
two effective chemical potentials produces new theoretical results for the 
spectral densities. In particular, non--Fermi liquid quasi--particles 
aquire a more complex structure (Eq. (\ref{poles})) than the 
case of equivalent planes.

\indent In order to calculate the carrier number per site per spin per 
plane, at $T = 0$, we have to perform the following integrals
\begin{equation}\label{123}
\rho_{1(2)} = \frac{1}{4}~\int^{+D}_{-D}~d\varepsilon~
\int_{-\omega_c}^{0}~N(\varepsilon) 
~A_{11(22)}(\varepsilon,\omega)~d\omega ~~~,
\end{equation}
\noindent which have to be computed numerically. Work is in 
progress to solve Eq. (\ref{123}) self--consistently. $N(\varepsilon)$ 
is the uncorrelated density of states. To conclude this Appendix, we
say that our results generalise the case of Yin and Chakravarty\cite{YCh}. 
With respect to the results of Ref.\cite{Hanke}, they have not discussed  
the two one--particle spectral functions, $A_{11}(\vec{k},\omega)$ 
and $A_{12}(\vec{k},\omega)$, around the effective chemical potential. 
Most likely the results of Ref.\cite{Hanke} are valid for large values of 
frequencies.

\newpage

\begin{figure}[h]
\centering
\includegraphics[clip,width=0.6\textwidth, height=0.35\textheight]{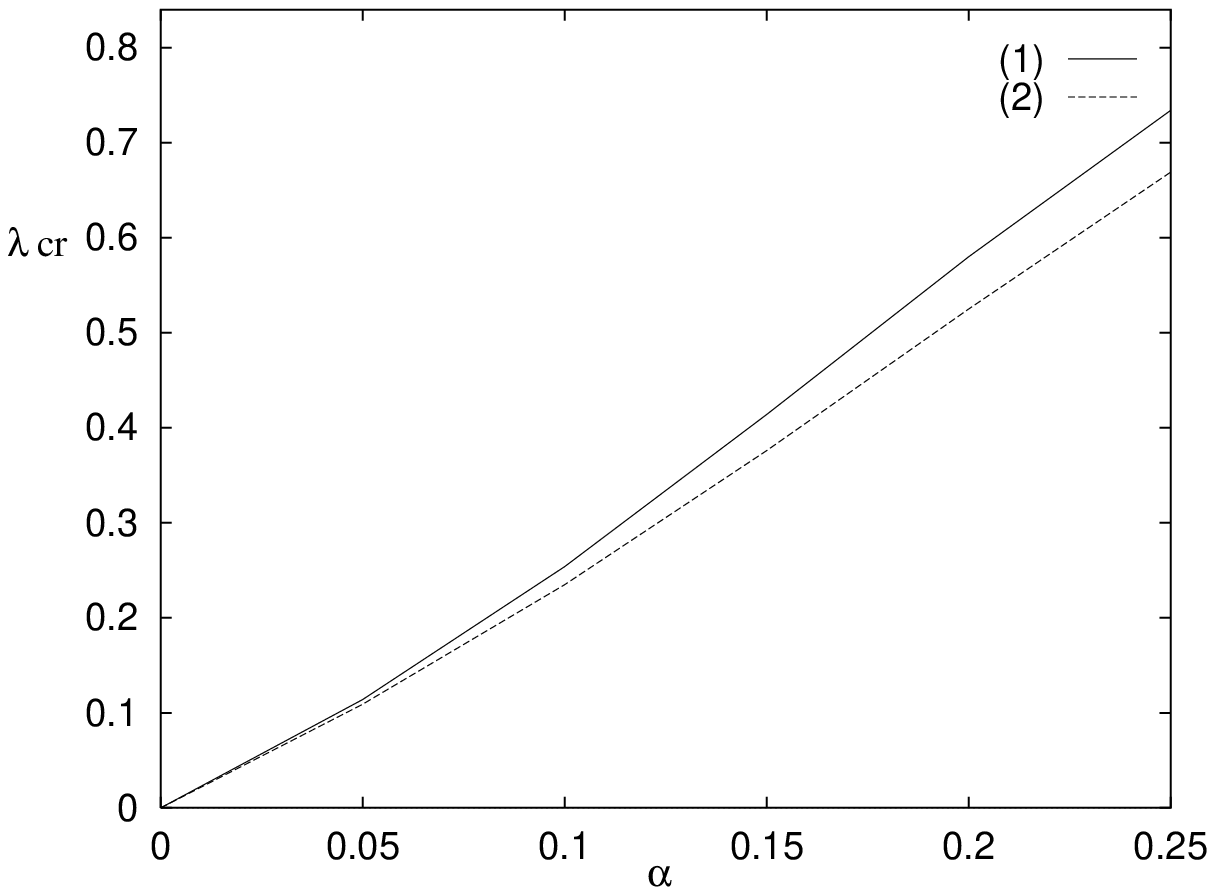}
\caption{ The critical coupling constant versus the 
non-Fermi parameter $\a$. The two lines correspond to the renormalized 
(1)-$\gamma=1/(1-\a$) and non-renormalized (2)-$\gamma=1$ case.}
\end{figure}

\begin{figure}[h]
\centering
\includegraphics[clip,width=0.6\textwidth, height=0.35\textheight]{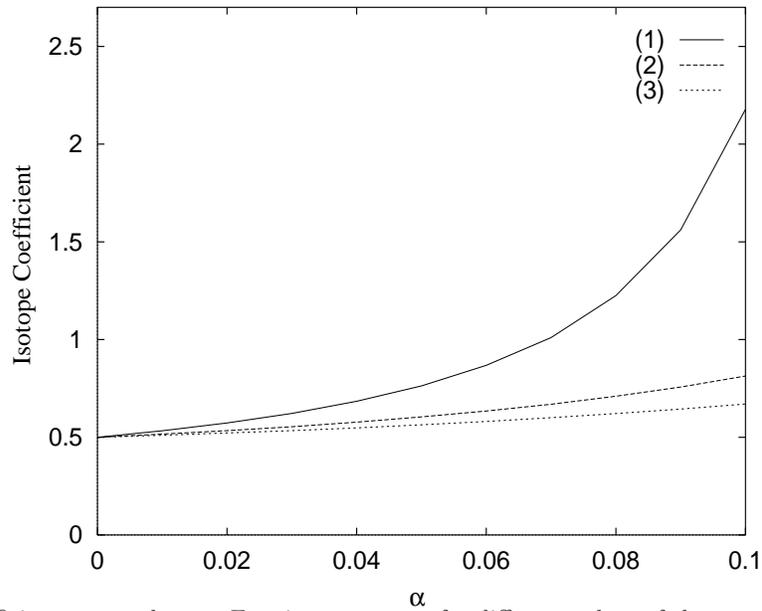}
\caption{The isotope coefficient versus the 
non-Fermi parameter $\a$ for different values of the coupling constant 
((1)-$\l=0.33$, (2)-$\l=0.66$. (3)-$\l=1$).}
\end{figure}

\newpage

\begin{figure}[h]
\centering
\includegraphics[clip,width=0.6\textwidth, height=0.35\textheight]{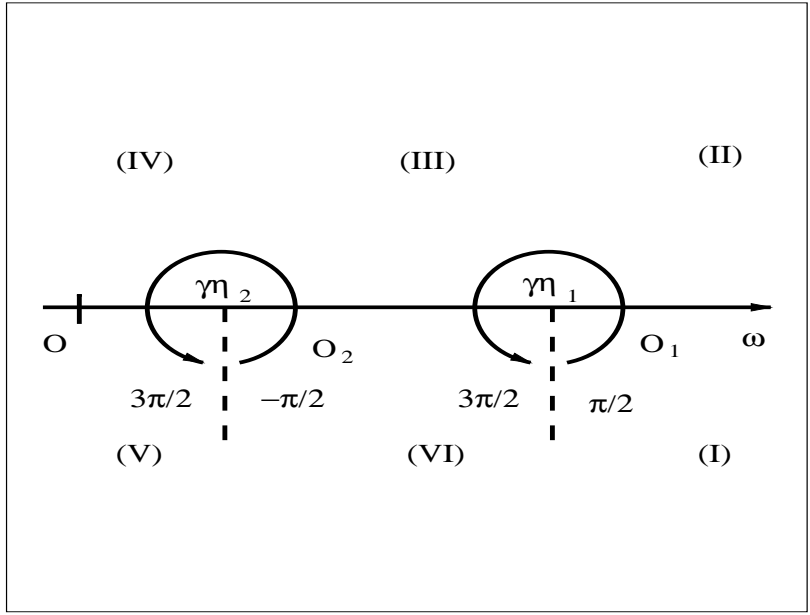}
$\vspace{1cm}$
\includegraphics[clip,width=0.6\textwidth, height=0.35\textheight]{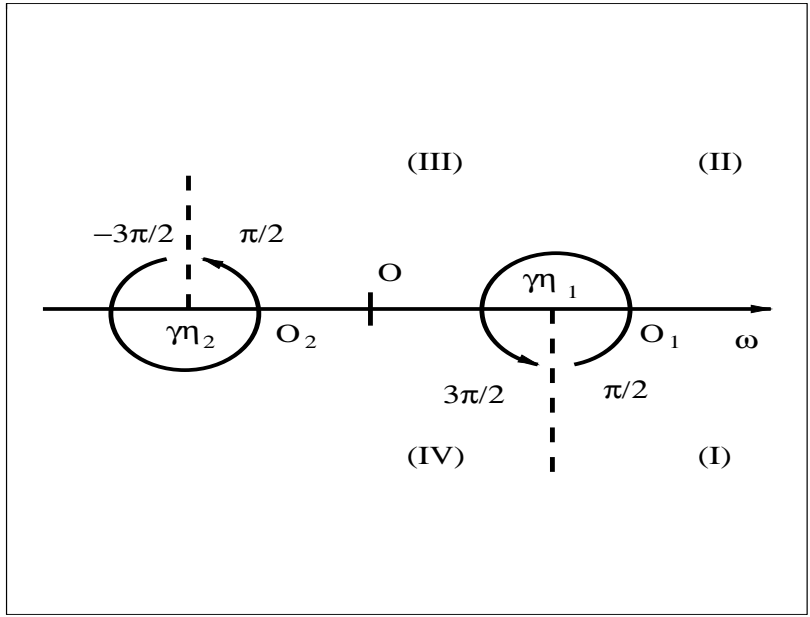}
$\vspace{1cm}$
\caption{(a)  For $\omega > 0$, $\gamma\eta_{1,\vec{k}} > 
\gamma\eta_{2,\vec{k}}$ and $\eta_{(1,2,)\vec{k}} > 0$, we have divided the 
complex plane in six regions, showing the branch points ($\gamma 
\eta_{2,\vec{k}}$ and $\gamma \eta_{2,\vec{k}}$) and branch 
cuts. The branch cuts are given by the dash lines. The solutions, 
for $\omega > 0$, are in regions: 1) $I$, namely, $n = 0$; and 2) $V$. 
(b) For $\omega > 0$, $\gamma\eta_{1,\vec{k}} >  
\gamma\eta_{2,\vec{k}}$, and $\eta_{1,\vec{k}} > 0$, 
$\eta_{2,\vec{k}} < 0$, we have divided the 
complex plane in four regions, showing the branch points and branch 
cuts. For $\omega > 0$ the solutions are in region $I$. }
\end{figure}

\newpage
\begin{figure}[h]
\centering
\includegraphics[clip,width=0.6\textwidth, height=0.35\textheight]{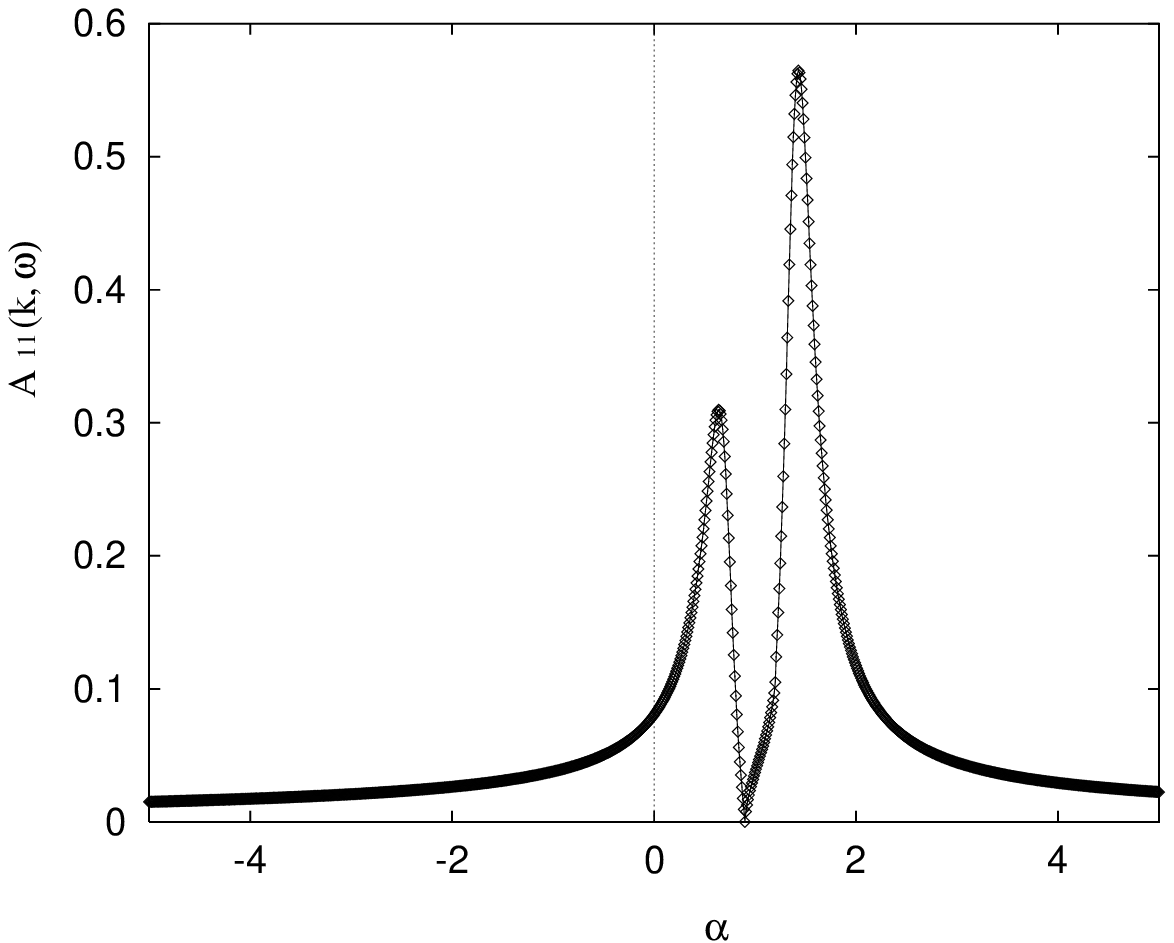}
$\vspace{1cm}$
\includegraphics[clip,width=0.6\textwidth, height=0.35\textheight]{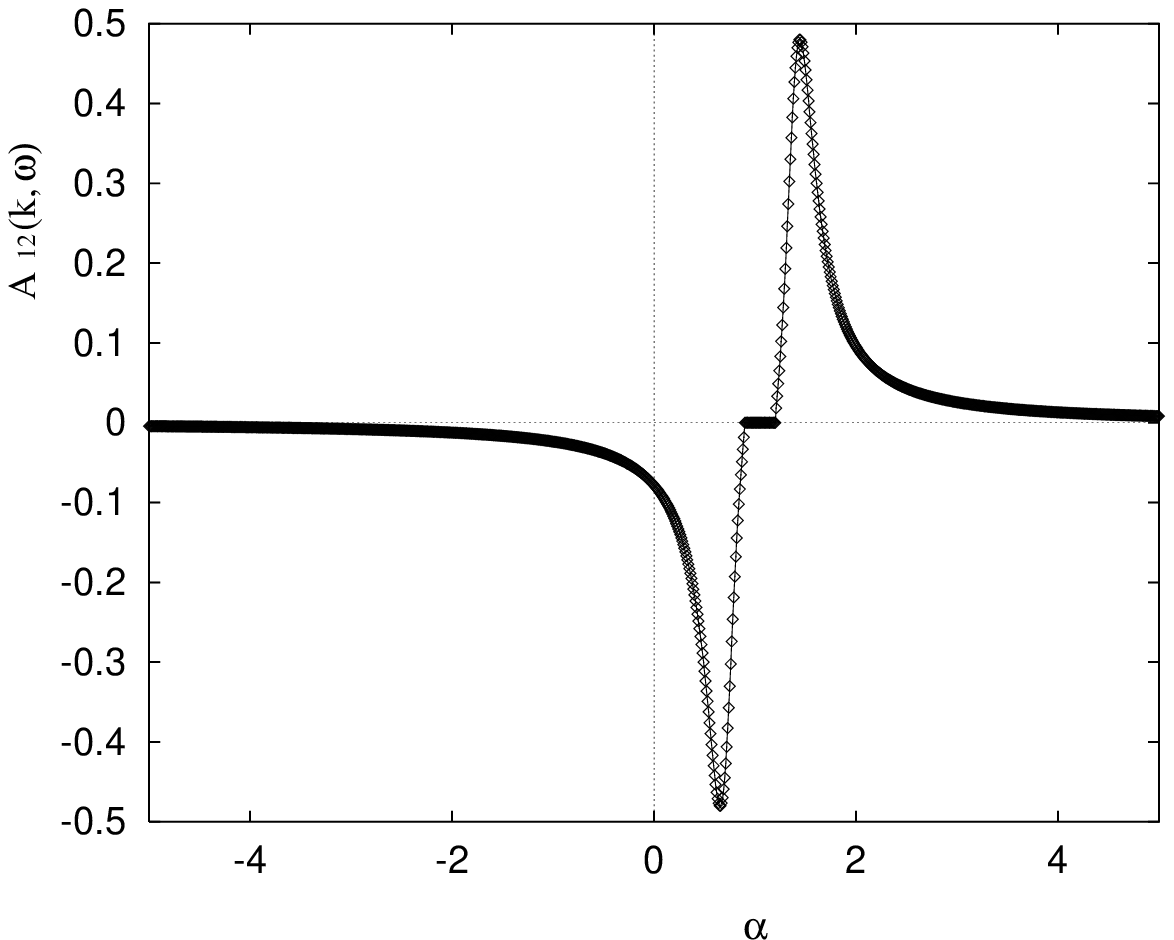}
$\vspace{1cm}$
\caption{(a) The diagonal one--particle spectral 
function, $A_{11}(\vec{k},\omega)$, vs $\omega$, for some 
values of $\gamma \eta_{1,\vec{k}}$ and $\gamma \eta_{2,\vec{k}}$,
namely, $\gamma \eta_{1,\vec{k}} = 1.2$ and 
$\gamma \eta_{2,\vec{k}} = 0.9$. Compare with $F_{11}(x)$ of 
Ref.[13]
We notice that the symmetry around $\omega = 0.9$ 
is lost. This is a realization of inequivalent 
coupled Hubbard layers. In similar form, 
$A_{22}(\vec{k},\omega)$, vs $\omega$ should not be 
symmetric around $\omega = 1.2$, according with the 
given parameters. (b)The off-diagonal one--particle spectral 
function, $A_{12}(\vec{k},\omega)$, vs $\omega$. Same values as 
in Figure 4.a. Compare with $F_{12}(x)$ of Ref.[13]. Now 
we have a gap in the energy spectrum betweem 
$\omega = 0.9$ and $\omega = 1.2$. This is also a consequence 
of having inequivalent coupled Hubbard layers.}
\end{figure}

\end{document}